\documentclass[sigconf]{acmart}

\AtBeginDocument{%
  \providecommand\BibTeX{{%
    \normalfont B\kern-0.5em{\scshape i\kern-0.25em b}\kern-0.8em\TeX}}}

\setcopyright{acmcopyright}
\copyrightyear{2022}
\acmYear{2022}
\setcopyright{acmlicensed}\acmConference[CHI '22]{CHI Conference on Human Factors in Computing Systems}{April 29-May 5, 2022}{New Orleans, LA, USA}
\acmBooktitle{CHI Conference on Human Factors in Computing Systems (CHI '22), April 29-May 5, 2022, New Orleans, LA, USA}
\acmPrice{15.00}
\acmDOI{10.1145/3491102.3501939}
\acmISBN{978-1-4503-9157-3/22/04}

\usepackage{listings}
\usepackage{subfig}
\usepackage{color, soul}
\usepackage[font={scriptsize,sf},tableposition=top]{caption}
\usepackage{float}
\floatstyle{plaintop}
\restylefloat{table}
\usepackage{multirow}
\usepackage{colortbl}
\usepackage{tabularx}
\usepackage{balance}
\usepackage[utf8]{inputenc}
\usepackage[T1]{fontenc}

\begin{document}

\title{Making Data Tangible: A Cross-disciplinary Design Space for Data Physicalization}

\author{S. Sandra Bae}
\email{sandra.bae@colorado.edu}
\affiliation{
  \department{ATLAS Institute}    
  \institution{University of Colorado}
  \city{Boulder, CO}
  \country{USA}
}

\author{Clement Zheng}
\email{clement.zheng@nus.edu.sg}
\affiliation{
  \department{Industrial Design}    
  \institution{National University of Singapore}
  \city{}
  \country{Singapore}
}
\author{Mary Etta West}
\email{mary.west@colorado.edu}
\affiliation{
  \department{Computer Science}    
  \institution{University of Colorado}
  \city{Boulder, CO}
  \country{USA}
}

\author{Ellen Yi-Luen Do}
\email{ellen.do@colorado.edu}
\affiliation{
  \department{ATLAS Institute}    
  \institution{University of Colorado}
  \city{Boulder, CO}
  \country{USA}
}
\author{Samuel Huron}
\email{samuel.huron@telecom-paristech.fr}
\affiliation{
  \department{Department SES - T\'{e}l\'{e}com Paris}    
  \institution{Institut Polytechnique de Paris, CNRS I3}
  \city{Palaiseau}
  \country{France}
}

\author{Danielle Albers Szafir}
\email{danielle.szafir@colorado.edu}
\affiliation{
  \department{ATLAS Institute}    
  \institution{University of Colorado}
  \city{Boulder, CO}
  \country{USA}
}
\affiliation{
  \department{\& Dept. of Computer Science} 
  \institution{University of North Carolina}
  \city{Chapel Hill, NC}
  \country{USA}
}

\renewcommand{\shortauthors}{Bae et al.}
\def\corpusNum{47}

\newcommand{\red}[1]{{\color{red} #1}}

\begin{abstract}
Designing a data physicalization requires a myriad of different considerations. Despite the cross-disciplinary nature of these considerations, research currently lacks a synthesis across the different communities data physicalization sits upon, including their approaches, theories, and even terminologies.
To bridge these communities synergistically, we present a design space that describes and analyzes physicalizations according to three facets: context (end-user considerations), structure (the physical structure of the artifact), and interactions (interactions with both the artifact and data). 
We construct this design space through a systematic review of 47 physicalizations and analyze the interrelationships of key factors when designing a physicalization. 
This design space cross-pollinates knowledge from relevant HCI communities, providing a cohesive overview of what designers should consider when creating a data physicalization while suggesting new design possibilities. 
We analyze the design decisions present in current physicalizations, discuss emerging trends, and identify underlying open challenges.
\end{abstract}




\maketitle
\section{Introduction}
\label{sec:introduction}
Data physicalization---the practice of mapping data to physical form---sits at the crossroads of various domains, including data visualization, tangible user interaction (TUI), and design\footnote{We specifically refer to the ``design community'' as the Design-oriented venues in CHI where researchers focus on broadening interaction design while promoting new aesthetic and sociocultural possibilities (i.e., the DIS conference and the CHI Design subcommittee). We acknowledge that researchers within the broader CHI community are all designers to some degree, but we use the distinction of the ``design community'' to emphasize the values that this specific group holds within HCI.}. Visualization researchers formally define physicalization as ``how computer-supported, physical representations of data (i.e., physicalizations) can support cognition, communication, learning, problem solving, and decision making'' \cite{jansen2015opportunities}. While physicalization is a nascent research subdiscipline, the fundamental ideas behind physicalization have a rich history that goes beyond visualization research. Examples of physicalizations have existed for centuries (e.g., Sumerian clay tokens~\cite{sumerian}, Ammaslik wooden maps~\cite{ammassalik},  Marshall Islands stick charts~\cite{marshall}) and are also integrated into crafting practices (e.g., steganography \cite{kuchera2018steganography}, computational textiles \cite{kurbak2018stitching}). As a result of this extensive and diverse history, data physicalization artifacts are diverse in structure, features, and interactions, appearing in many different contexts.

Crafting physicalizations requires design considerations beyond conventional visualization approaches. While all visualizations must consider the expressivity of their designs (e.g., data encoding and data interactions), data physicalizations must also be mindful of structural considerations (e.g., structural stability, material choice) and the context where the physicalization exists (e.g., audience, location).
The diversity and interplay of these requirements can make physically representing data difficult.
Data physicalization designers currently face an additional burden of having to navigate different relevant intellectual communities' set of approaches, theories, and terminologies without a cohesive lens. As a result, designers may be unaware of the knowledge and best practices emerging in other communities such as data visualization (i.e., how to effectively encode data), TUI (i.e., how to amplify the capabilities of the human body and physical world for interaction design), or design (i.e., how to identify and leverage a material’s potential). However, bridging diverse knowledge and practice requires understanding each community's values, biases, and sets of assumptions. 
To holistically understand the depth and breadth of the design dimensions data physicalization provides, this paper examines physicalizations from the perspective of researchers from the data visualization, TUI, and design communities. 

While previous research has cataloged physicalization artifacts based on semiotics \cite{zhao-embodiment2008, vandemoere-physicality, offenhuber2015indexical}, fabrication techniques \cite{djavaherpour2021rendering}, and visualization tasks \cite{dragicevic:hal-02113248}, past work often focuses on a single perspective of data physicalization rather than reflecting on the broader intellectual foundations of the communities in HCI that data physicalization sits upon.
We instead aim to synthesize knowledge from across these communities to develop a theoretical design space that highlights cross-disciplinary considerations for creating physicalizations.
While design spaces can be characterized by their (1) descriptive power, ``to describe a significant range of existing interfaces''; (2) evaluative power, ``to help assess multiple design alternatives'', and (3) generative power, ``to help designers create new designs'' \cite{Beaudouin2004}, this design space is more \textit{descriptive} by nature to focus on the aforementioned aim. We develop this design space by examining key components of physicalization design and development: who, what, where, and how. \textit{Who} are the intended audiences (e.g., the public, scientists, analysts)?  \textit{Where} are physicalizations used or showcased? \textit{What} design decisions contribute to a physicalization's materiality (i.e., presence in the world)? \textit{What} interactions does a physicalization afford? \textit{How} do different communities approach physicalization? (And what can we learn from each other?) These questions help elucidate and describe specific factors of physicalization design and their interrelationships to highlight the emerging patterns in physicalization research.

\begin{figure*}[tb]
    \centering
    \includegraphics[width=\textwidth]{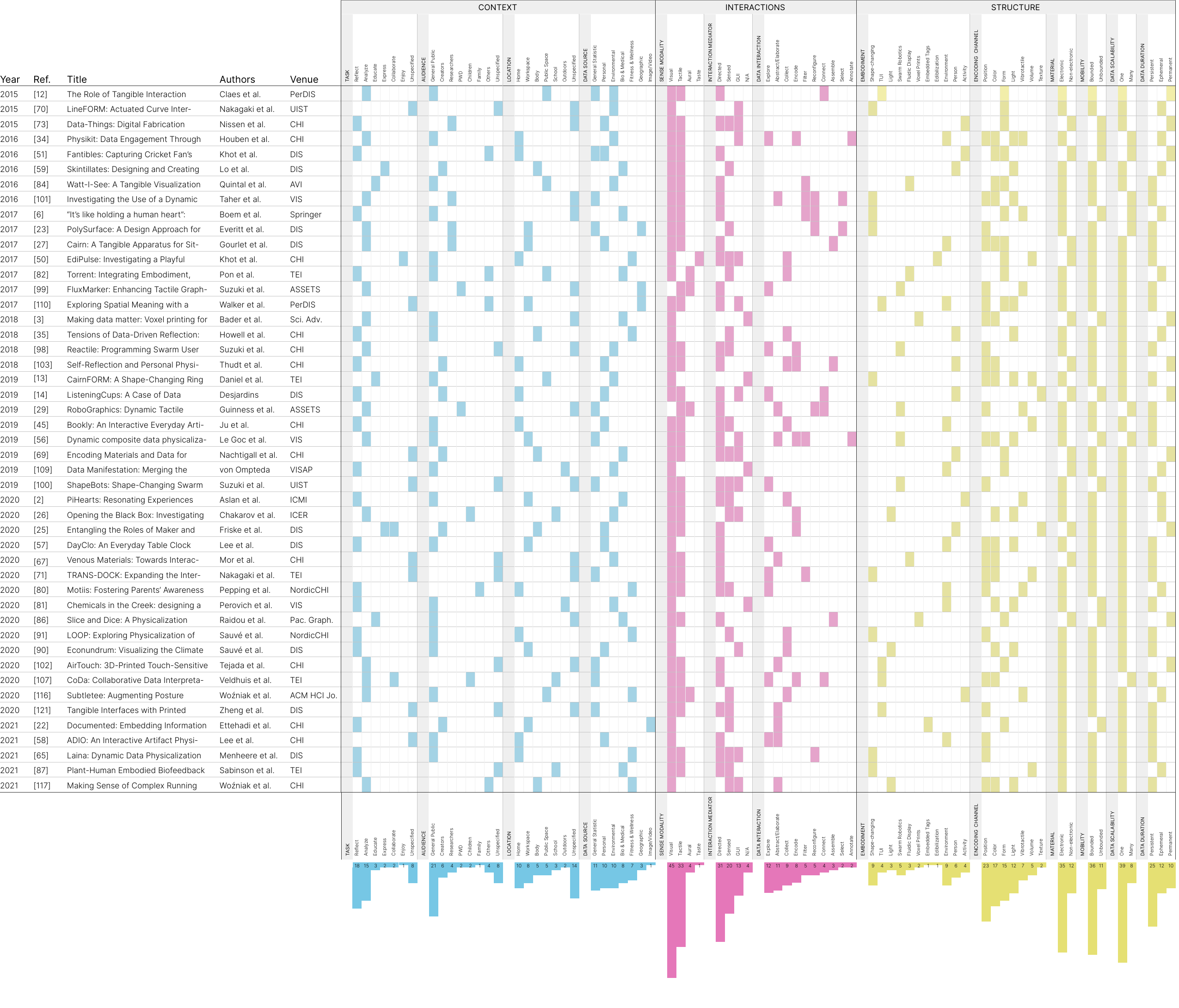}
    \caption{We surveyed physicalizations from across HCI to construct a design space spanning three facets---context (end-user considerations),  structure (physical structure of the artifact), and interactions (interactions with both the artifact and data)---over 13 dimensions. While dimensions primarily reside in one facet, each design dimension is not siloed to just one facet but rather can influence other facets: the design dimensions are interrelated. See Section~\ref{sec:codebook} for more details. The bar on the outer frames shows the distribution of the design dimension within our surveyed corpus. The length of each bar is determined by how many times the design dimension has been used.}
    \label{fig:framework}
    \Description{Conceptual Framework for Data Physicalization. This conceptual framework picture is a matrix that reflects the coding process. The matrix contains 3 main columns, labelled as: Context, Interactions, and Structure. The Context column contains Task, Audience, Location and Data Source. The Interaction column contains Sense Modality, Interaction Mediator, and Data Interaction. The Structure column contains Data Duration, Data Scalability, Mobility, Material, Encoding Channel, and Embodiment. The bottom of each column has bars of different lengths for each item. Task has long bars for Reflect, Analyze, and short bars for Educate, Express, Collaborate, Enjoy, and a medium long bar for Unspecified. Audience has a long bar for General Public, and short bars for Creators, Researchers, PWD, Children, Family, Other, and a medium bar for Unspecified. Location has medium bars for Home, Workspace, short bars for Body, Public Space, School, Outdoors, and Medium bar for Unspecified. Data Structure has medium bars for General Statistics, Personal, Environmental, Bio and Medical, Fitness and Wellness, and short bars for Geographic, and Image/Video. Sense Modality has really long bars for Vis, Tactile, and short bars for Aural and Taste. Interaction Mediator has long bar for Directed, medium bar for Sensed and GUI, and a short bar for N/A. Data Interaction has medium bars for Explore, Abstract/Elaborate, Collect, Encode, and short bars for Filter, Connect, Reconfigure, Assemble, Select and Annotate. Data Duration has medium bars for Persistent, Ephemeral, and Permanent. Data Scalability has a long bar for One, and a short bar for Many. Mobility has a long bar for Bounded, and a short bar for unbounded. Material has a long bar for Electronic and a short bar for Non-Electronic. Encoding Channel has medium bars for Position, Color, Form, Light, and short bars for Vibrotactile, Volume, and Texture. Embodiment has short bars for Shape-Changing, Swarm Robotics, Light, TUI, Fluidic Display, Embedded Tags, Voxel Prints, Ediblization, Environment, Person and Activity.}
\end{figure*}

We developed our design space by surveying physicalization research in two popular databases encompassing relevant communities: the ACM Digital Library (DL) and IEEE Xplore.
Our survey aims to capture the rapidly emerging practices of physicalization research since Jansen et al.'s formalization of the field in 2015 \cite{jansen2015opportunities}. We look at work that has been published after this date to examine advances since their initial theoretical effort and limit our scope to recent advances. Our survey used a grounded theory-inspired approach to analyze various interdisciplinary design considerations in \corpusNum\space physicalizations from 18 different venues. 
We surveyed these physicalizations according to three facets of design---context (the end-user considerations), structure (the physical structure of the object), and interactions (interactions with both the artifact and data)---deriving thirteen dimensions of physicalization design, illustrated in Figure~\ref{fig:framework}. We developed our codes by drawing on the diverse experience of the authors to help characterize and identify key considerations and open challenges in design, computation, interaction, and materiality \cite{vallgarda2007computational, wiberg2010computational, vermeulen2013cross, offenhuber2021physicality}. Our design space bridges the knowledge from each community: it provides a cohesive overview of what designers should consider when creating a data physicalization and summarizes relevant trade-offs. We close by highlighting promising opportunities for different communities to learn from each other and offering guidance for future work in the field.

\begin{figure*}[t]
  \centering
  \includegraphics[width=0.65\textwidth]{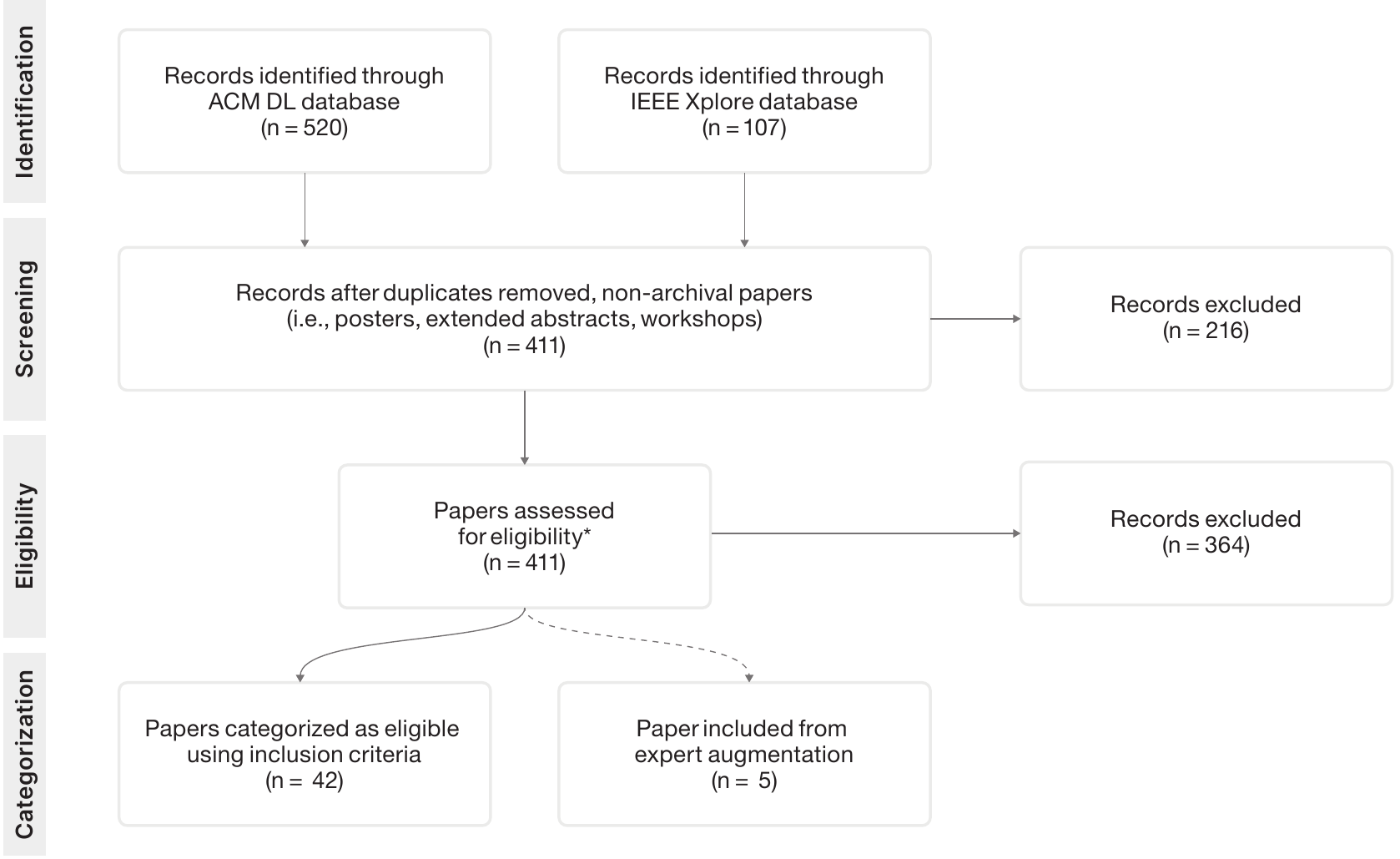}
  \caption{PRISMA flow diagram \cite{page2020prisma} on how the final corpus was curated ($n=47$). See Section~\ref{sec:dataset} for our inclusion criteria.}
  \label{fig:prisma}
  \Description{Flow diagram showing corpus curation process. Left side has 4 labels, from top to bottom are Identification, Screening, Eligibility, and Categorization. The row of Identification has two boxes, the left box shows the text Records identified through ACM DL database (n = 520), and the right box shows the text Records identified through IEEE Xplore database (n = 107). Both boxes have down pointing arrows to the next box at the Screening row, that’s labelled Records after duplicates removed, non-archival papers (i.e., posters, extended abstracts, workshops) (n = 411). This box has a right pointing arrow to a box that’s labelled Records excluded (n = 216), and a downward arrow to the box in the Eligibility row, labeled Papers assessed for eligibility* (n = 411). This box has a right pointing arrow to a box that shows Records excluded (n = 364). The Paper assessed for eligibility* (N = 411) has two downward pointing arrows to the Categorization row, left solid arrow to a box that is labeled Papers categorized as eligible using inclusion criteria (n = 42), and a dashed downward pointing arrow to the box that’s labelled Paper included from expert augmentation (n = 5)}
\end{figure*}

\section{Background \& Related Work}
\label{sec:related}
To create a design space for physicalization, we draw on existing definitions of the field and explore past characterizations of physicalization from visualization, TUI, and design.

\subsection{What is Data Physicalization?}
\label{sec:definition}
The definition and practice of data physicalization are constantly evolving. Jansen et al. \cite{jansen2015opportunities} originally defined physicalization as ``a physical artifact whose geometry or material properties encode data'' \cite{jansen2015opportunities}. This definition implies that only the physical properties of the artifact encode data. However, Hogan and Hornecker \cite{hogan2012multimodal} extend this definition by arguing that data in physicalizations can be ``encoded in the shape as well as the behavior of the representation.'' 
Using a physicalization made from spin tops as an example, Hogan and Hornecker contend that our interpretation of the data embedded in the spin tops cannot be fully interpreted until the spin top is set in motion. This example illustrates that our understanding of physicalization must go beyond the physical form and encompass a holistic view that considers the object's materiality (i.e., the quality of presence in the world \cite{gross2014structures, wiberg2018materiality}), a well-studied concept in design research but less common in visualization.
Offenhuber \cite{offenhuber2021physicality}, in contrast, contends our understanding of data physicalization should be based on broadening our traditional notion of data (i.e., external data sets) rather than simply focusing on the physical artifact. 
By employing epistemological and ontological perspectives, Offenhuber illustrates how many physicalization examples are not based on external datasets and highlights our implicit and explicit assumptions about data, especially what we consider data to be.
These perspectives illustrate how the definition and values behind data physicalization are still in flux. 
We, however, do not attempt to define data physicalization. Instead, we acknowledge 
that the definition of physicalization varies among and even within the communities. This variance necessitates cross-disciplinary synthesis to scaffold collaboration and innovation. 
 
Physicalization designers may view and implement physicalizations according to the practices and values of their native intellectual community. As one example, Friske et al. from the design community reflect on how during the process of creating their physicalization ``[d]ata became one material among others, no more or less important in design'' \cite{friske2020data}. With this perspective, Friske et al. discussed how they had to negotiate between how accurately they wanted to encode the data in the physical representation versus embracing the spirit of the material (i.e., yarn).
This perspective stands in contrast to the visualization community's more cognitively-oriented perspective where data encoding is typically prioritized in terms of its effectiveness in helping users carry out their analytical tasks, such as comparing values or estimating correlation \cite{card1999readings, munzner2015visualization}. 
Jansen et al. \cite{jansen2015opportunities} outlined the need to understand the effectiveness of physical variables (e.g., how do people perceive sizes in a physicalization \cite{jansen2015psychophysical}) as part of physicalization's research agenda. 
Offenhuber~\cite{offenhuber2021physicality} cautions that framing physicalization as another traditional visualization application may be a ``lost opportunity'' that unnecessarily restricts the conceptual and aesthetic implications of physicalization. We reference these perspectives throughout the paper in the context of trends in physicalization research to understand how different communities approach data. 

Our goal is to create a cohesive, cross-disciplinary survey focused on the design practices and decisions that researchers engage with when creating physicalizations. This survey considers an artifact as a physicalization if Jansen et al.'s or Hogan and Hornecker's definitions are satisfied.
Jansen et al.'s definition encompasses many of the accepted definitions in current research (see Section~\ref{sec:codebook} for our inclusion criteria); however, we augment this definition with Hogan and Hornecker's inclusion (i.e., data can also be encoded through the behavior of the artifact) to increase the diversity of sampled artifacts, including soft wearables ~\cite{howell2018tensions, friske2020data}, fluidic displays ~\cite{mor2020venous, pon2017torrent, quintal2016watt}, shape-changing displays ~\cite{taher2016emerge, everitt2017polysurface, nakagaki2020transdock, boem2018s}, and vibrotactile haptic displays~\cite{aslan2020pihearts, pepping2020motiis, wozniak2020golf} (Figure~\ref{fig:embodiment}). The resulting corpus provides an opportunity to reflect on different theoretical and design perspectives across disciplines.

\subsection{Physicalization Theory}
To develop broad, cross-disciplinary knowledge, we draw on existing theoretical investigations of physicalization. 
For example, Dragicevic et al. \cite{dragicevic:hal-02113248} outline historical and recent physicalizations organized by Brehmer's and Munzner's typology of abstract tasks \cite{brehmer2013multi}: to discover, to present, and to enjoy. This survey highlights examples of physicalizations that illustrate each task but does not consider other attributes of physicalizations that affect their use and comprehension, such as materiality. However, this task-centric perspective primarily reflects the core values of the visualization research community (i.e., designing systems that help people gain specific information from data \cite{sedlmair2012design}). Further, 
Dragecevic et al.'s survey focuses more on providing a landscape of artifacts as opposed to a systematic analysis of design factors such as material and location. 

Other efforts catalog and describe physicalization by focusing on specific design capabilities and epistemological perspectives, such as indexicality \cite{zhao-embodiment2008, offenhuber2015indexical}, autographical interactions between data and representation \cite{offenhuber2020autographic}, embeddedness  \cite{willet2017embedded}, situatedness \cite{bressa2021s}, and multi-sensory capabilities \cite{hogan2012multimodal}. These works illuminate the importance of a single critical aspect of physicalization design rather than reflecting on the broad intellectual foundations of the field. More practice-based investigations have explored the interdisciplinary interchange between visualization and design-based fields. For example, Hull and Willett \cite{hull2017architecture} explore how architectural methods can inform how designers can best engage an audience, encode data into physical form, and determine what kind of workflows can best support data physicalization. Hull and Willet's work highlights the promise of bridging cross-disciplinary methods to benefit physicalization. 

While past work illustrates the promise of physicalization from a range of perspectives, 
we lack design frameworks that accurately and holistically characterize this emerging research space. We take inspiration from Hull and Willet's work to bridge communities in a way that produces synergistic knowledge of the range of factors people consider in physicalization development, the interrelations of these factors, and key trade-offs. We build on prior theoretical efforts to formulate our preliminary set of codes (see Section~\ref{sec:codebook}) and expand on these ideas to illustrate how the knowledge from data visualization, TUI, and design can enrich our understanding of data physicalizations. We focus our survey on HCI research papers and exclude practice-based physicalizations from art and commercial tools. This focus allows us to maintain a manageable scope and minimize assumptions about intention, context, and design deliberations and trade-offs by using concrete, explicit evidence from the papers (e.g., descriptions, implementation details, design specifications) to analyze the author(s)' intentions. In contrast, artistic and public installations often provide minimal descriptions of the intentions and aims behind the works, requiring us to infer design intention and specifications and increasing the potential for errors in our coding process.

\section{Survey Methodology}
\label{sec:methodology}
To construct a design space for data physicalizations, we conducted a systematic survey based on a keyword search to assemble a corpus of physicalizations and augmented our corpus with canonical examples. These examples draw from the diverse experience of the authors, who are experts in visualization, TUI, and design. The survey data was then classified through an open coding approach to identify key design space dimensions. This review qualitatively analyzes work over a six-year period (2015--2021, $n=\corpusNum$; Figure~\ref{fig:prisma}) across visualization, TUI, and design.
Two authors served as coders throughout the process, working in regular consultation with the rest of the research team.

A keyword-based approach (the keywords are explained in Section~\ref{sec:dataset}) provides two benefits: (i) it imparts a clear cut-off criterion for papers published under the theme and/or reference of data physicalization and (ii) it can lead to more comprehensive results by efficiently sampling a larger range of possible papers.
However, different communities may use different terminology, meaning that keyword use may overlook niche terms of art. The result is also subject to the databases' primary research focus which can fail to encapsulate how communities beyond ACM and IEEE view data physicalization. To address these limitations, we supplement our keyword search with selected case studies drawn from the collective experience of the research team (five total additional studies). This expert augmentation identifies canonical examples based on the authors' knowledge of the field, providing critical depth to the design space beyond the initial search.

\subsection{Dataset Creation}
\label{sec:dataset}
The broad list of communities in which data physicalization research is published prohibited a systematic review of specific conference proceedings. Instead, we selected archival papers using keywords across ACM DL and IEEE Xplore---two of the most popular databases for data physicalization research that
encompass a range of venues (ACM CHI, ACM UIST, ACM DIS, IEEE VIS, etc.).
The resulting corpus is available in the supplemental materials\footnote{See \url{https://sandrabae.github.io/dataphys-gallery/}}.

Our theoretical framework is based on a corpus of \corpusNum\space physicalizations selected using the following inclusion criteria:\vspace{6pt}

\noindent\textbf{Date}: Though examples of physicalizations date prior to 2015, Jansen et al. coined the term ``data physicalization'' and formalized the field within visualization in 2015. We look at work that has been published after this date to examine advances since their initial theoretical effort and to limit the scope to recent advances. 
\vspace{6pt}

\noindent\textbf{Keyword Match}: We considered papers where the authors used one of the following terms in the title, abstract, text, or author keywords: \texttt{data physi-} (e.g., \texttt{data physicalization}, \texttt{physicalisation}, \texttt{data physicality}), \texttt{data materiality}, \texttt{tangible-} (e.g., \texttt{tangible data}, \texttt{tangible visualization}, \texttt{tangible visualisation}), and \texttt{physical-} (e.g., \texttt{physicalization}, \texttt{physical visualization}). We chose these terms due to their close relationship to data physicalization (e.g., physical visualization is a synonym for data physicalization \cite{jansen2013physical}) and their use in prior literature. 
\vspace{6pt}

\noindent\textbf{Eligibility}: Our initial search retrieved 627 candidate papers: 520 papers from ACM DL and 107 papers from IEEE Xplore. We then screened the candidate papers to remove duplicates\footnote{When we encountered the same work by the same authors, we used the latest published work (e.g., Le Goc et al., 2018 \cite{le2018dynamic}) and discarded the earlier versions from the corpus (e.g., Le Goc et al., 2016 \cite{legoc2016zooids}).} and non-archival papers (i.e., posters, extended abstracts, magazine articles, workshop papers). After screening, 411 papers remained. We then removed any papers that prohibited conclusive coding using the following criteria: 
\begin{enumerate}
    \item The paper introduced a new technique, briefly using one of the keywords but not explaining how the application or design of the technique can be applied to physicalization (e.g., \cite{braley2018grid, sahoo2016joled}).
    \item The artifact created in the paper did not provide data insights but was predominantly used as an interface control (e.g., \cite{manshaei2019tangible, Kaninsky2018bats, jofre2015tangible}). See the discussion of ``Data Physicalization and TUI'' in Jansen et al. \cite{jansen2015opportunities} for further discussion. 
    \item The paper presented an experiment focused on outlining empirical guidelines (e.g., \cite{jansen2016size, Stusak2016mind, garcia2021scale}). We prioritized papers that demonstrated a design process through the artifact produced in the paper. 
    \item Data was encoded through sonification or olfaction rather than through the physical properties of an object (e.g., \cite{nath2020sonifcation, batch2020olfactory}). Sonification \cite{madhyastha1995data, kaper1999data} and olfaction \cite{marco2012signal} are considered their own areas of research and outside of our current scope.
    \item Papers that presented multiple artifacts \cite{lockton2020sleepecologies, karyda2020narrative}, but the individual artifacts presented ambiguous codes.
    To reduce ambiguity, we prioritized papers that allowed insight into the overall design methodology used to develop an artifact.
\end{enumerate}

Using these inclusion criteria, two research team members each examined half of the corpus to remove any papers that did not meet the criteria or needed further discussion. Papers needing discussion were reviewed by the full research team. After deriving the final dataset, we added five additional examples~\cite{boem2018s, bader2018making, vonompteda2019sea, raidou2020slice, Nachtigall2019encoding} based on recommendations from the team, who reflect expertise in visualization, TUI, and design, to capture relevant examples not included in the original corpus. 

\begin{table*}[t]
  \small
  \centering
  \begin{tabular}{p{0.0625in}>{\raggedright}p{1.2in}>{\raggedright}p{3.2in}l}
    \toprule
    \multicolumn{2}{p{1.5625in}}{\textbf{Design Dimension}} &
    \textbf{Code} &
    \textbf{Multiple}\\
    \arrayrulecolor{black!30}\midrule

    \multicolumn{2}{p{1.2625in}}{\textbf{Context}} & & \\

    &
    Task &
    Analyze, Educate, Express, Reflect, Enjoy, Collaborate, Unspecified & Yes\\
    \addlinespace[0.2em]
    &
    Audience &
    General Public, Children, Family, Researchers, People with Disability, Creator, Other, Unspecified &
    Yes \\
    \addlinespace[0.2em]
    &
    Location &
    Home, Workspace, Public Space, Body, Outdoors, School, Unspecified & No \\
    \addlinespace[0.2em]
    &
    Data source &
    Biological and Medical, General Statistics, Personal, Fitness \& Wellness, Geographical, Environmental, Image/Video &  Yes \\
    
    \midrule
    
    \multicolumn{2}{p{1.2625in}}{\textbf{Structure}} & & \\
    &
    Embodiment &
    Shape-Changing Display, Swarm Robotics, Fluidic Displays, Voxel Prints, Light Display, Embedded Tags, Edibilization, TUI, Environmentally-Embedded, Personally-Embedded, Activity-Embedded & No \\
    \addlinespace[0.3em]
    &
    Material &
    Non-Electronic, Electronic  & No \\
    \addlinespace[0.2em]
    &
    Encoding Channel &
    Color, Form, Position, Volume, Texture, Light, Vibrotactile  & Yes \\
    \addlinespace[0.2em]
    &
    Mobility &
    Unbounded, Bounded  & No \\
    \addlinespace[0.2em]
    &
    Data Scalability &
    One, Many & No \\
    \addlinespace[0.2em]
    &
    Data Duration &
    Ephemeral, Persistent, Permanent & No \\
    \midrule
    
    \multicolumn{2}{p{1.2625in}}{\textbf{Interactions}} & & \\

    &
    Interactions Mediator &
    GUI, Direct, Sensed, N/A  & Yes \\
    \addlinespace[0.2em]
    &
    Sense Modality &
    Visual, Tactile, Aural, Taste & Yes \\
    \addlinespace[0.2em]
    &
    Data Interactions &
    Select, Explore, Reconfigure, Encode, Abstract/Elaborate, Filter, Connect, Annotate, Collect, Assemble & Yes \\
    \arrayrulecolor{black}\bottomrule
  \end{tabular}
  \caption{The final codebook with 13 design dimensions and 74 subcodes. Dimensions were further categorized into those describing the \textit{context}, \textit{structure}, and \textit{interactions} of a physicalization. \textit{Multiple} refers to situations where multiple codes may describe a single dimension according to the paper's images and supporting text. Full code definitions can be found in the supplemental materials.}
  \label{tab:codebook}
\end{table*}

\subsection{Codebook}
\label{sec:codebook}
We coded each example in our corpus using a grounded theory-inspired approach. Our codebook consisted of terms and categories that examined the three key facets of physicalization design and development (context, structure, and interaction), focusing on five questions: \textit{Who} are the intended audiences?  \textit{Where} are physicalizations used or showcased? \textit{What} design decisions contribute to a physicalization's materiality? \textit{What} interactions does a physicalization afford? \textit{How} do different communities approach physicalization? (And what can we learn from each other?) We generated terms and categories that described the variation within our corpus guided by these questions and past theory from relevant communities (e.g., using interaction models from data visualization \cite{yi2007interactions} and design \cite{vermeulen2013cross}), adding terms as we observed new design variations that did not fit our initial codes. These codes are intended to reveal the interrelationships between key factors of data physicalization by drawing upon existing theories from the relevant communities.
We limited our codes to attributes of the physicalization that could be assessed through the paper's images and supporting text. This constraint precluded analysis where we would have to make assumptions about the authors' intention or approach. The final codebook is summarized in Table~\ref{tab:codebook} and full code definitions can be found in the supplemental materials.

We summarize the 13 dimensions of our design space by whether they pertain to a physicalization's \textit{context} (i.e., aspects of the design environment such as users, data, or location), its \textit{structure} (i.e., its physical form), and its \textit{interactions} (i.e., how it affords and responds to user inputs).

\subsubsection{Context}
\label{sec:context} Context describes the physical, social, and analytical frames a physicalization addresses \cite{chalmers2004historical}. Context grounds the visual design and functionality behind a data physicalization, including the attributes describing the intended goals, environment, and users of the physicalization. We use the concept of \textit{situated visualization}---embedding data into its context of use---from the visualization community as a lens to see how physicalizations make use of context. Context also examines the end-user considerations guiding design, including its intended \textit{tasks} (what is the functional purpose of the physicalization), \textit{audience} (who is the physicalization intended for), \textit{location} (where does the physicalization exist), and \textit{data source} (what kinds of data are represented). 

\subsubsection{Structure}
\label{sec:structure}
Structure focuses on how data is physically embodied and encoded, including its overall form. We draw on theoretical perspectives from the TUI and design communities on embodiment and materiality to characterize structure. Categorizing physicalizations based on the physical structure is challenging, largely due to the diverse fabrication approaches that physicalization designers adopt \cite{djavaherpour2021rendering}. We focus on how the data is \textit{embodied}---how data is physically manifested \footnote{Our use of \emph{embodiment} refers to providing a tangible or visible form to something abstract \cite{pahl1996embodiment}. This definition is distinct from embodiment as considering human bodies and actions in tangible and social computing systems, as described by Dourish \cite{dourish1999embodied}.}. The concept of embodiment and its benefits have been extensively discussed in the TUI community \cite{ishii1997tangiblebits, hornecker2005design, wyeth2008engagement}. For example, Zhao \& Vande Moere \cite{zhao-embodiment2008} who also recognize the importance of embodiment for physicalizations. 

To holistically understand embodiment in physicalization, we also examine the physicalization's use of materials, such as 3D printing filament, yarn, motors, and liquids. Materials are the building blocks of physicalizations, shaping a physicalization's physical structure, interactions, and materiality (i.e., presence in the world \cite{gross2014structures, wiberg2018materiality}).
To understand the relationship between materials and other dimensions of physicalization design, we coded examples of what \textit{materials} the physicalization is composed of; how data is \textit{encoded} through the physicalization's tangible and visual properties; how \textit{mobile} the physicalization is; whether the physicalization is \textit{scalable} in handling multiple, distinct datasets; and how long the data is displayed for (i.e., \textit{data duration}). Through these design dimensions, we can also gain a better understanding of the values designers place in the material. Gross et al. \cite{gross2014structures} state there are three distinct perspectives a designer can consider when using a material: physical, computational, or cultural. The physical framing focuses on how a material functions with its physical properties and how effectively a material can accomplish a task. The computational framing denotes that computers, digital information, and other intangible aspects of computational systems are also considered materials that are incorporated with physical things. The cultural aspect focuses on examining materials related to crafting practices and how materials ``communicate'' to the maker. 

\subsubsection{Interactions}
\label{sec:interactions-codebook}
Interaction focuses on the dialogue between the user and the physicalization. With physicalizations, a \textit{data interaction} (e.g., filtering a dataset) can be examined by analyzing how the interaction was executed at the artifact-level (i.e., what design feature in the artifact made it possible to execute the filter interaction?). We draw on the concepts of affordances, feedforward, and feedback \cite{vermeulen2013cross} to examine how interactions in general (e.g., pressing a button) are possible at the artifact-level, and we use Yi et al.'s  data interaction framework \cite{yi2007interactions} to identify the different data interactions imposed on physicalizations.

We characterize interactions based on how the physicalization affords interactions and what actions are taken with the data.
The \textit{interactions mediator} refers to how an individual's data interactions are captured (e.g., through a direct action, use of GUI, sensed through a sensor). The \textit{sensing modality} highlights which human senses are used when receiving feedback.
\textit{Data interactions} examine how each interaction changes the data presented to the user. 
We use Yi et al.’s framework of data interactions  \cite{yi2007interactions} to help capture the user’s intent while interacting with the physicalization. We added three additional interactions (annotate, collect, assemble) based on observations from our corpus that incorporate actions that are more natural in physical rather than digital systems. 
\textit{Annotating} refers to adding supplemental information to the physicalization, \textit{collecting} provides additional data to the physicalization, and \textit{assembling} focuses on piecing physicalizations together.

\subsection{Coding Process \& Agreement}
\label{sec:coding_process}
Two coders independently coded the physicalizations using a preliminary set of design dimensions. They reconciled codes after the first 12 papers to ensure common understanding and application and again after each new code was added.
The inter-rater reliability of the preliminary codes using Krippendorff's alpha \cite{krippendorff2018content} was high ($\kappa=0.85$; 95\% agreement). The coders discussed any conflicting codes with the rest of the research team to arrive at a consensus, with one set of final codes assigned to each physicalization. 

\section{Design Space}
\label{sec:summary}
The corpus consists of a total of \corpusNum\space papers. This corpus covers several conferences in HCI and Design, including CHI (14/47), DIS (10/47), TEI (5/47), UIST (2/47), ASSETS (2/47), NordicCHI (2/47), PerDIS (2/47), AVI (1/47), ICER (1/47), and ICMI (1/47). We also survey conferences from Visualization and Visual Analytics (4/\corpusNum) \footnote{IEEE VIS (2/47), IEEE VISAP (1/47)} and journal articles (3/\corpusNum).\footnote{The Proceedings of the ACM on Human Computer Interaction (HCI), Springer AI \& Society, and Science Advances} From our coding process, we derived 13 design dimensions that describe critical decisions in physicalization, which we organized into three facets: context, structure, and interactions (Section~\ref{sec:design-space-context} - Section~\ref{sec:interactions}). 
We use these three facets to discuss the five questions we posed in Section~\ref{sec:introduction}: \textit{Who} are the intended audiences?  \textit{Where} are physicalizations used or showcased? \textit{What} design decisions contribute to a physicalization's materiality? \textit{What} interactions does a physicalization afford? \textit{How} do different communities approach physicalization? (And what can we learn from each other?). See Figure~\ref{fig:framework} and  \url{https://sandrabae.github.io/dataphys-gallery/} to see how the 13 design dimensions are distributed throughout the corpus.

\subsection{Context}
\label{sec:design-space-context}
\noindent\textbf{Tasks}: While the data visualization community typically designs for exploration \cite{card1999readings, munzner2015visualization}, we found \textit{reflection}---the act of increasing one's self-knowledge---to be the most popular task physicalizations support (18/\corpusNum).  Topics of reflection varied with focuses including personal informatics (e.g., exercise \cite{khot2016fantibles, sauve2020loop}, reading habits \cite{ju2019bookly}, emotions \cite{howell2018tensions}) and data about family (e.g., children's emotions \cite{pepping2020motiis}). People also used physicalizations as a way to make a personal connection to bigger communities, such as commemorating a local river flood catastrophe \cite{pon2017torrent}, permit violations in a local river \cite{perovich2021chemicals}, and climate change \cite{vonompteda2019sea}. 

The next most common task was \textit{analysis} (15/\corpusNum). In these cases, people used physicalizations to explore statistical datasets. Physicalizations were also used to \textit{educate} (3/\corpusNum), \textit{collaborate} (2/\corpusNum), \textit{express} (2/\corpusNum; i.e., using a physicalization to express their feelings or ideas), and \textit{enjoy} (1/\corpusNum). There were also a handful of instances (8/\corpusNum) where the task was \textit{unspecified}. In these cases, physicalizations were introduced but their intended purpose and/or what users should gain from the artifact was not explicit. 

\noindent\textbf{Audience.} We found that the majority of physicalizations targeted the \textit{general public} (21/\corpusNum), focusing on incorporating data into everyday activities \cite{nissen2015data,howell2018tensions, boem2018s, ju2019bookly} that required little visualization literacy. Beyond the general public, we found physicalizations designed with a specific audience in mind, including \textit{creators} (6/\corpusNum) (i.e., the physicalization is meant for the person who created it), \textit{researchers} (4/\corpusNum), \textit{children} (2/\corpusNum), \textit{people with disabilities} (2/\corpusNum), and \textit{family} (1/\corpusNum). We found four instances of \textit{other} (4/\corpusNum) designed for musicians \cite{pon2017torrent}, golfers \cite{wozniak2020golf}, sports fans \cite{khot2016fantibles}, and college campus visitors \cite{walker2017spatial}. Several papers (8/\corpusNum) did not explicitly state the intended audience of the artifact. In these cases, the physicalizations were designed to showcase a novel technology where they did not further explain their intended audience. While many physicalizations are often created with one audience group in mind, Torrent \cite{pon2017torrent} (Figure~\ref{fig:time}a) is an interactive system that augments concert flute music with live water sounds designed for both performers and their audience throughout a performance. Water surface turbulence encodes the  flutists’ muscle tension, and the physicalization is situated between the performer and audience to provide information throughout the performance. 

\noindent\textbf{Location}. 
Physicalizations in the corpus were designed for a range of locations, such as workspaces, homes, public spaces, outdoors, and even on the body. Many of the physicalizations were meant for a private location, including \textit{homes} (10/\corpusNum), the \textit{body} (5/\corpusNum), and \textit{workspaces} (8/\corpusNum). About a quarter of the papers focused within a \textit{public space} (8/\corpusNum), such as \textit{schools} \cite{chakarov2020black, veldhuis2020coda}, golf ranges \cite{wozniak2020golf}, and event venues \cite{pon2017torrent, daniel2019cairn}. Two physicalizations specifically focused on \textit{natural environments} (ocean \cite{vonompteda2019sea} and river \cite{perovich2021chemicals}). Approximately a third of the papers in the corpus (14/\corpusNum) provided little to no information about the physicalization's intended location. As with the physicalization's audience, papers with an \textit{unspecified location} often used physicalization as part of an illustrative example for a new technique.

Physicalizations most commonly resided in homes (10/\corpusNum). These physicalizations focused on capturing data from the house (e.g., environmental data \cite{houben2015physikit}, children's emotions during gaming \cite{pepping2020motiis}) or mimicking everyday objects, such as clocks \cite{lee2020dayclo}, lamps \cite{ju2019bookly}, mementos \cite{khot2016fantibles}, and design pieces \cite{sauve2020loop, menhree2021laina, lee2021adio}. Physicalizations for the home were designed to seamlessly integrate into the home environment. Everyday objects designed and situated for private environments, such as the home, encouraged more private and personally meaningful reflections \cite{pepping2020motiis, ju2019bookly, sauve2020loop}. Laina \cite{menhree2021laina}, for example, is a 2.5D shape-changing display designed as an art piece that represented the user's running route data. Laina's slow-motion design features allowed the users to reflect on their running path and feel a sense of pride in their accomplishments. Similarly, people who used Bookly \cite{ju2019bookly} mentioned that Bookly's design, which encouraged readers to place and pick up the book at the base of the lamp opened to their current progress, prompted them to read more and be more reflective about their reading habit.

\noindent\textbf{Data Source.} 
Compared to traditional visualization \cite{sarikaya2018design}, few physicalizations focused on traditional static, tabular data (11/\corpusNum).
Target tabular datasets included the U.S. population \cite{suzuki2019shapebots}, world GDP \cite{taher2016emerge, nakagaki2020transdock}, civic data \cite{claes2015tangible}, conference metadata \cite{tejada2020airtouch}, student’s grades \cite{le2018dynamic}, and sports matches \cite{khot2016fantibles}. 
Several physicalizations supported \textit{biomedical and medical} data (8/\corpusNum), primarily using bio-sensing data, such as galvanic-skin response \cite{howell2018tensions}, gait \cite{Nachtigall2019encoding}, heart rate \cite{boem2018s,lo2016skintillates,khot2017chocolate}, respiration rate \cite{Sabinson20121plant}, and muscle tension \cite{pon2017torrent}. \textit{Fitness and wellness} datasets (7/\corpusNum) also included bio-sensing data, but overall aimed to look at the data for more of a wellness perspective, in terms of improving their physical ability \cite{wozniak2021runningshoes, wozniak2020golf}, setting exercising goals \cite{khot2016fantibles, sauve2020loop}, and bringing more awareness of their physical activity \cite{menhree2021laina} and emotions \cite{howell2018tensions}. Physicalizations also frequently supported reflection on other forms of \textit{personal} data (10/\corpusNum), such as one's reading habit \cite{ju2019bookly}, schedule \cite{lee2020dayclo}, diet \cite{sauve2020econumdrum}, fabrication lab practices \cite{gourlet2017cairn}, and Twitter data \cite{nissen2015data,khot2016fantibles}. Finally, there are physicalizations that focus on encoding \textit{environmental} data (10/\corpusNum), touching upon topics such as energy \cite{daniel2019cairn, quintal2016watt, nakagaki2015line}, weather \cite{suzuki2018reactile}, and climate change \cite{vonompteda2019sea}, and \textit{geographic} data (3/\corpusNum). Documented \cite{ettehadi2021documented} was the only artifact that used an \textit{image/video} dataset (1/\corpusNum). 

\vspace{-1.3em}
\subsection{Structure}
\label{sec:space_structure}
\textit{Shape-changing} interfaces were the most popular way physicalizations embodied data (9/\corpusNum). Other forms of physical embodiment included \textit{swarm robotics} (5/\corpusNum), \textit{TUIs} (4/\corpusNum), \textit{fluidic displays} (3/\corpusNum), \textit{light displays} (3/\corpusNum), \textit{voxel prints} (2/\corpusNum), \textit{embedded tags} (1/\corpusNum), and \textit{edibilization} (1/\corpusNum). Data was also physically embodied through \textit{environments} (9/\corpusNum), \textit{people} (6/\corpusNum), and \textit{activities} (4/\corpusNum). See Figure~\ref{fig:embodiment} for an overview of physical embodiments. 
Several physicalizations (13/\corpusNum) attempted to co-opt existing idioms behind common visualizations (e.g., line chart, bar chart, scatterplot). However, others used a custom data representation, leveraging the unique affordances of physicalization (see Section~\ref{sec:agency} for more details). 

The majority of surveyed physicalizations incorporated \textit{electronics} as part of their structure (35/\corpusNum), while only a quarter of the physicalizations (12/\corpusNum) did not contain any form of electronics (i.e., \textit{non-electronic}). Though the majority of the physicalizations incorporated various solid materials (43/\corpusNum), five physicalizations incorporated liquids as part of their main material choice \cite{mor2020venous, pon2017torrent, quintal2016watt,perovich2021chemicals,vonompteda2019sea}. Liquid displays leveraged a number of physical attributes to encode data, including microfluidics  \cite{mor2020venous}, use of vortexes as metaphors for energy production \cite{quintal2016watt}, surface turbulence in a pool reflecting muscle use during a musical performance \cite{pon2017torrent}, and as the location for situated data \cite{vonompteda2019sea, perovich2021chemicals}. We also observed a single instance where the researcher embraced computational materiality \cite{vallgarda2007computational}---the perspective that computers, digital information, and other intangible aspects of computational systems are also considered to be materials that can be incorporated with a physical object. Documented \cite{ettehadi2021documented} contains an embossed visual tag that was designed to be embedded within a 3D printed object. To support DIY makers, it focuses on retrieving documentation images and videos using a smart device where the digital information contributes to the 3D object's materiality.

We observed a connection between the use of materials and whether data is represented ephemerally (i.e., data is displayed for a short period of time), persistently (i.e., data is displayed for as long as the user desires), or permanently (i.e., data is permanently embedded into the physical structure of the physicalization). Approximately half of the physicalizations were able to display data \textit{persistently} (25/\corpusNum) through programmatic means, while 12 displayed data \textit{ephemerally} (12/\corpusNum) by streaming data from sensors to outputs such as LEDs \cite{lo2016skintillates, chakarov2020black}, vibrotactile motors \cite{aslan2020pihearts, pepping2020motiis}, and pneumatics \cite{Sabinson20121plant} in real-time. Ten artifacts displayed data \textit{permanently} with non-electronic materials (10/\corpusNum). Despite how long data was displayed, \textit{position} was the most popular data encoding method (23/\corpusNum), followed by \textit{color} (17/\corpusNum), \textit{form} (15/\corpusNum), use of \textit{light} (12/\corpusNum), \textit{vibrotactile} (7/\corpusNum), \textit{volume} (5/\corpusNum), and \textit{texture} (2/\corpusNum). 

\subsection{Interactions}
\label{sec:interactions}
The surveyed physicalizations supported 61 total interactions. These interactions were categorized using the seven data interaction categories from Yi et al., \cite{yi2007interactions} as well as annotate, collect, assemble (see Section~\ref{sec:interactions-codebook}). Of the ten data interactions, \textit{explore} was the most pervasive (12/61), followed by \textit{abstract/elaborate} (11/61), \textit{collect} (9/61), \textit{encode} (8/61), \textit{filter} (5/61),
\textit{reconfigure} (5/61),
\textit{connect} (4/61),  \textit{assemble} (3/61),  \textit{select} (2/61), and
\textit{annotate} (2/61). The majority of physicalizations enabled \textit{direct interactions} (i.e., introducing interactions by requiring users to intentionally touch or manipulate the representation of data; 41/61), while some relied only on \textit{sensors} (9/61) or required users to interact with a \textit{GUI} (7/61).
\section{Findings}
\label{sec:findings}
Section~\ref{sec:summary} summarizes key distributions of different design decisions across the 13 design dimensions. 
For brevity, we now use those findings to discuss patterns in the design space that help bridge dimensions or disciplines; reveal how structure, context, and interactions are interrelated to each other; and offer compelling future opportunities. 

\subsection{Who are the intended audiences?}
\label{sec:audience}
Physicalizations commonly targeted the general public (21/\corpusNum), focusing on how to incorporate data into everyday activities \cite{nissen2015data,howell2018tensions, boem2018s, ju2019bookly} that required little visualization literacy. Our observation matches Jansen et al. \cite{jansen2015opportunities}'s description of the benefits of data physicalization---active perception and engagement with the data---as physicalizations commonly target accessible ambient knowledge and awareness of pervasive personal or community issues. However, people's intended engagement varied throughout our corpus. For example, Physikit \cite{houben2015physikit}, a system of four cubes with environmental sensors embedded inside each cube, provides active engagement. Each cube outputs the sensor data (e.g., noise, humidity, temperature) either through light, vibration, movement, or airflow. Due to its small size and discrete nature, people actively placed the Physikit cubes throughout their homes to compare environmental sensor data in the different parts of their house. Bookly \cite{ju2019bookly}, in contrast, focuses on more ambient engagement.
Bookly is an interactive lamp that tracks an individual's accumulated reading time in addition to other reading metrics.
Bookly's design seamlessly integrates reading progress data through a mountain-shaped holder. The mountain-shaped holder provides an intuitive place where users should place their book and emphasizes that the reading is still ongoing by keeping the pages open. Bookly's overall form enables the system to naturally collect data by tracking a user's reading time once the book is picked up.
Through its volumetric disk, Bookly ambiently displays the user's accumulated reading time with respect to their reading goal while the pages of the open book communicate progress. 

Beyond the general public, several physicalizations were designed with a specific audience in mind, including \textit{researchers} (4/\corpusNum), \textit{children} (2/\corpusNum), and \textit{people with disabilities} (2/\corpusNum). Physicalizations meant for these specific sets of users focused on data exploration tasks. However, the depth of the supported data analysis varied depending on the interactions the physicalizations supported. Zooids \cite{le2018dynamic} and EMERGE \cite{taher2016emerge}, for example, provide analytical insight from the rich number of data interactions both systems provides. Zooids is a swarm robotic system where users are able to map data instances (e.g., a student) with multiple attributes (e.g., grades, gender, letter recommendation score) to each robot. The system provides users the ability to \textit{encode}, \textit{abstract/elaborate}, \textit{annotate}, and \textit{filter} the resulting multi-dimensional data. 
Based on the combination of these data interactions, Zooids support  collaborative decision-making, such as faculty deciding on student admissions or a family planning a vacation. 
EMERGE \cite{taher2016emerge} similarly integrated \textit{filter, reconfigure, abstract/elaborate}, and \textit{select} into their 2.5D shape-changing display where users present and share analytical insight from a dataset. Robographics \cite{guinness2019robographics} and FluxMarkers \cite{suzuki2017flux} encourage data exploration for users with disabilities, specifically addressing the needs of blind and low vision (BLV) people. However, Robographics and FluxMarkers supported a limited range of interactions. Both systems supported only elaborating data details (e.g., the minimum/maximum value of an x-axis) by combining auditory and physical interactions. 

We also found six instances where the physicalization was meant for the \textit{creators} themselves. Often, these physicalizations are in the domain of personal informatics (i.e., personally relevant information for the purpose of self-reflection and self-monitoring \cite{elsden2015personal}). In the process of creating the physicalizations, creators tended to use custom data mappings to incorporate meaningful reflections (see Section~\ref{sec:summary} to see the topics of reflections illustrated in the corpus). When analyzing and exploring the physicalization, only the creators would have insights on what the data is about. ListeningCups \cite{desjardins2019Listening}, for example, are tactile cups where each bump and groove on the surface of the cup represents the everyday ambient noise within the authors' environment. Aside from the authors who created the artifacts, others would not have an immediate context as to what data has been encoded. The authors in contrast described how they gained personal insights from their data when they held the cups during meandering conversations or while washing the dishes. This led to reflections of self-inquiries such as, ``\textit{Why is the part feeling more smooth, why were we so quiet?}''. 
Friske et al. \cite{friske2020data} explore tensions between physicalizations use by the creator and by external audiences. They argue that a personal informatics physicalization does not have one ``true'' data interpretation.
They question the authority of data, noting how no one has the authority to invalidate someone's data interpretation even if they did not create the artifact. They elaborate that these interpretations should not be seen as distinct but as informing each other and the ongoing understanding of the data itself.

\begin{figure*}[t]
  \centering
  \includegraphics[width=\textwidth]{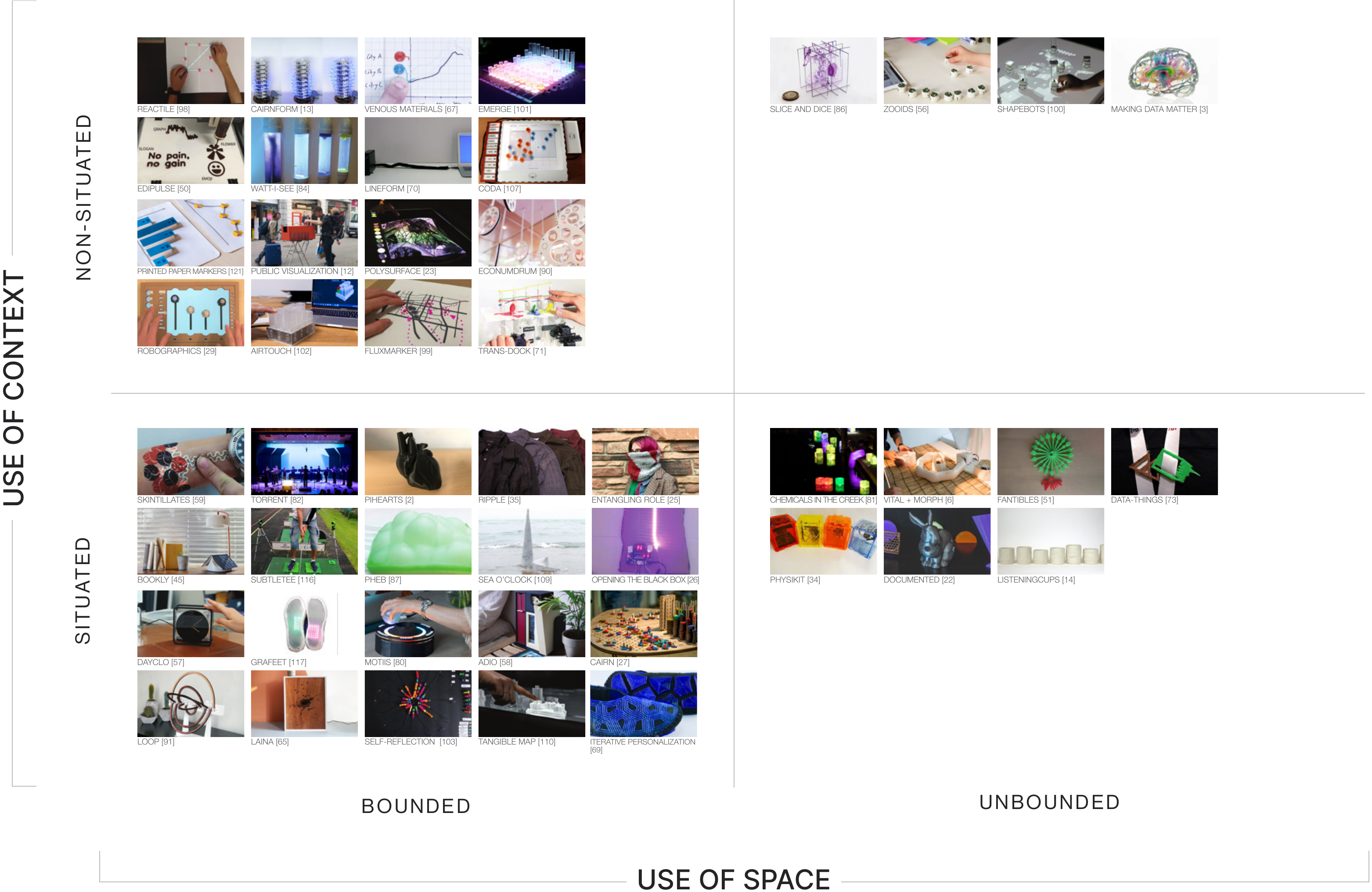}
  \caption{Physicalizations classified according to their use of space and their use of context. \textit{Bounded} artifacts are confined to a particular position and not meant to be repositioned. \textit{Unbounded} artifacts are designed for mobility and to move freely across space. \textit{Situated} and \textit{non-situated} refer whether the physicalization is displaying information that is contextually relevant in its environment. All images are copyright to their respective owners.}
  \label{fig:bounded}
  \Description{Classifications of physicalizations. A cross in the middle divided this figure into 4 quadrants. The left label for the horizontal axis is labelled as User of Context, the top if Non-situated, and the bottom Situated. The bottom label for the vertical axis is labelled Use of Context. The left is labelled Bounded, and the right Unbounded. The top left quadrant (Non-situated, Bounded) contains 16 square images, each one represents a paper. The top right quadrant (Non-situated, Unbounded) contains 4 square project images.The bottom left quadrant (Situated, Bounded) contains 20 square project images. The bottom right quadrant (Situated, Unbounded) contains 7 square project images.}
\end{figure*}

\subsection{Where are physicalizations used or showcased?}
\label{sec:location}
Analyzing how physicalizations use space provides insight into how the physicalization leverages its environment. 
For example, a physicalization's situatedness---a concept within the visualization \cite{elsayed2015situated} and other relevant HCI communities (e.g., Ubiquitous Computing \cite{schilit1994context, weiser1991computer})---describes how information is integrated into its surrounding environment. However, the relationship of the physicalization to its location can also change: physicalizations are either \textit{bounded} or \textit{unbounded} to a given location depending on their mobility (Section~\ref{fig:bounded}). Bounded artifacts exist confined to a particular place and are not meant to be re-positioned. Unbounded artifacts are designed for mobility and may be relocated by users or natural forces, such as a river flow (e.g., Perovich et al. \cite{perovich2021chemicals}). Figure~\ref{fig:bounded} shows the distribution of the physicalizations using these two concepts.

Traditional notions of situatedness from visualization focus on how a specific physical location provides context to the presented data to support data comprehension and decision making. While physicalizations are often situated by location (e.g., \cite{vonompteda2019sea, perovich2021chemicals}), we found that situatedness also occurs through individuals (i.e., \textit{body-embedded} physicalizations where the physicalization is integrated with the body) and with respect to activities (i.e., \textit{activity-embedded} physicalizations where the data reflects a captured or current state of an activity, as discussed in Bressa et al. \cite{bressa2021s}). 

Body-embedded situatedness includes physicalizations that are meant to be worn on the body as either clothing \cite{howell2018tensions, friske2020data}, shoes \cite{Nachtigall2019encoding}, accessories \cite{wozniak2020golf}, or tattoos \cite{lo2016skintillates}. In many of these cases, the data originates from the wearer through external sensors (e.g., \cite{howell2018tensions, lo2016skintillates}). However, Friske et al. \cite{friske2020data} encode data through knitting, but the knitter is still the physical referrent for the data. Similarly, Wo\'{z}niak et al. \cite{wozniak2020golf} used arm and leg bands 
to provide vibrotactile feedback of the wearer's elbow bend and body weight imbalance to improve their golfing stance. The arm band is situated with respect to the golfer but the situatedness extends beyond the wearer and into the activity itself, dynamically reacting to the swing in the context of the golfer's motions. These physicalizations expand the traditional notion that situatedness is siloed to location. 

The majority of physicalizations are bound to one location (36/\corpusNum). Unbounded physicalizations are generally composed of discrete parts that can freely move across space, such as ShapeBots \cite{suzuki2019shapebots}, Zooids \cite{le2018dynamic}, Vital + Morph \cite{boem2018s}, and Chemicals in the Creek \cite{perovich2021chemicals}. For example, Vital + Morph is a shape-changing display allowing families to co-monitor their loved one's vital signs. Vital + Morph can be disassembled into smaller pieces where each family member can monitor different vital signs at various locations within a city. As an unbounded physicalization, Vital + Morph can be moved freely within and between  environments across space. Being distributed across
the different family members brings a sense of community and creates a shared understanding, illustrating how physicalizations can be situated beyond their physical environment. These examples reflect and expand on the perspectives of Bressa et al. \cite{bressa2021s}, who state situatedness can occur across space, time, place, activity, and through communities. 
We additionally found that physicalizations can be situated through the individual (i.e., the body).  

\begin{figure}[b]
  \centering
  \includegraphics[width=\columnwidth]{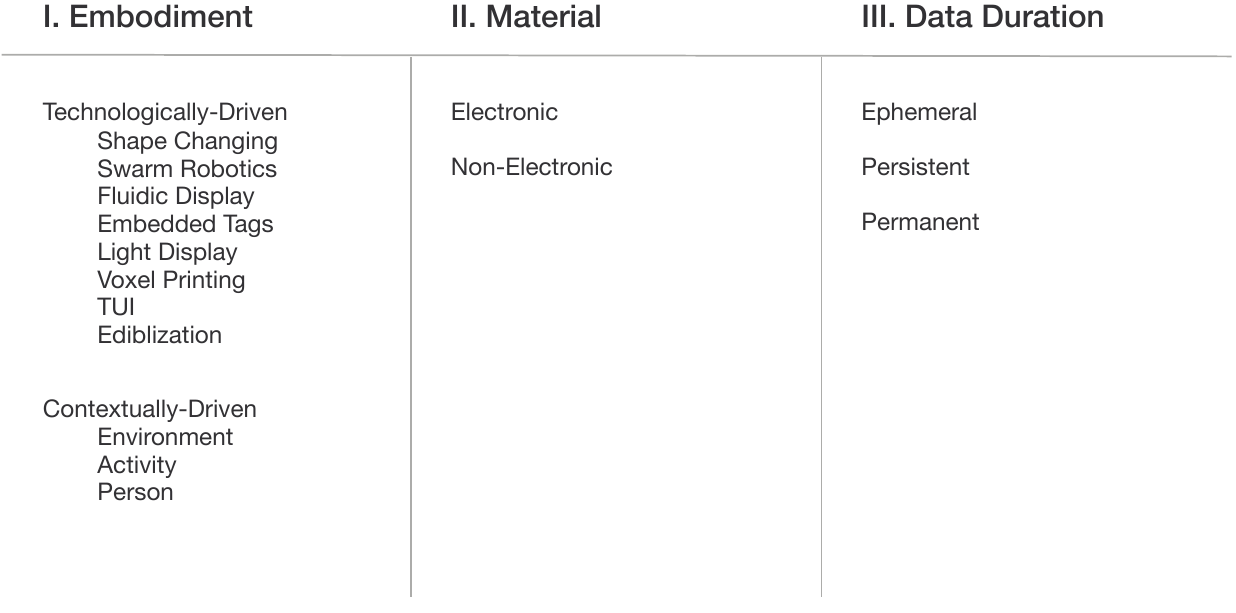}
  \caption{Conceptual space that describes the three attributes that contribute to a physicalization's materiality (i.e., presence in the world \cite{gross2014structures, wiberg2018materiality}).}
  \label{fig:materiality}
  \Description{Conceptual space of the Physicalization Materiality. It has 3 columns. First column on the left is labelled Data Duration. Below it has 3 items, Persistent, Ephemeral, and Permanent. Second column is labelled Embodiment. Below it shows 2 subcategories. First is Technologically-Driven, that includes items Shape-Changing, Embedded Tags, TUI, Swarm Robotics, Light Display, Edibilization, Fluidic Display, and Voxel Prints. Second is Contextually-Driven, that includes Environment, and Activity. The Third column is labelled Material. Below it are Electronic, and Non-electronic.}
\end{figure} 

\begin{figure*}[t]
  \centering
  \includegraphics[width=\textwidth]{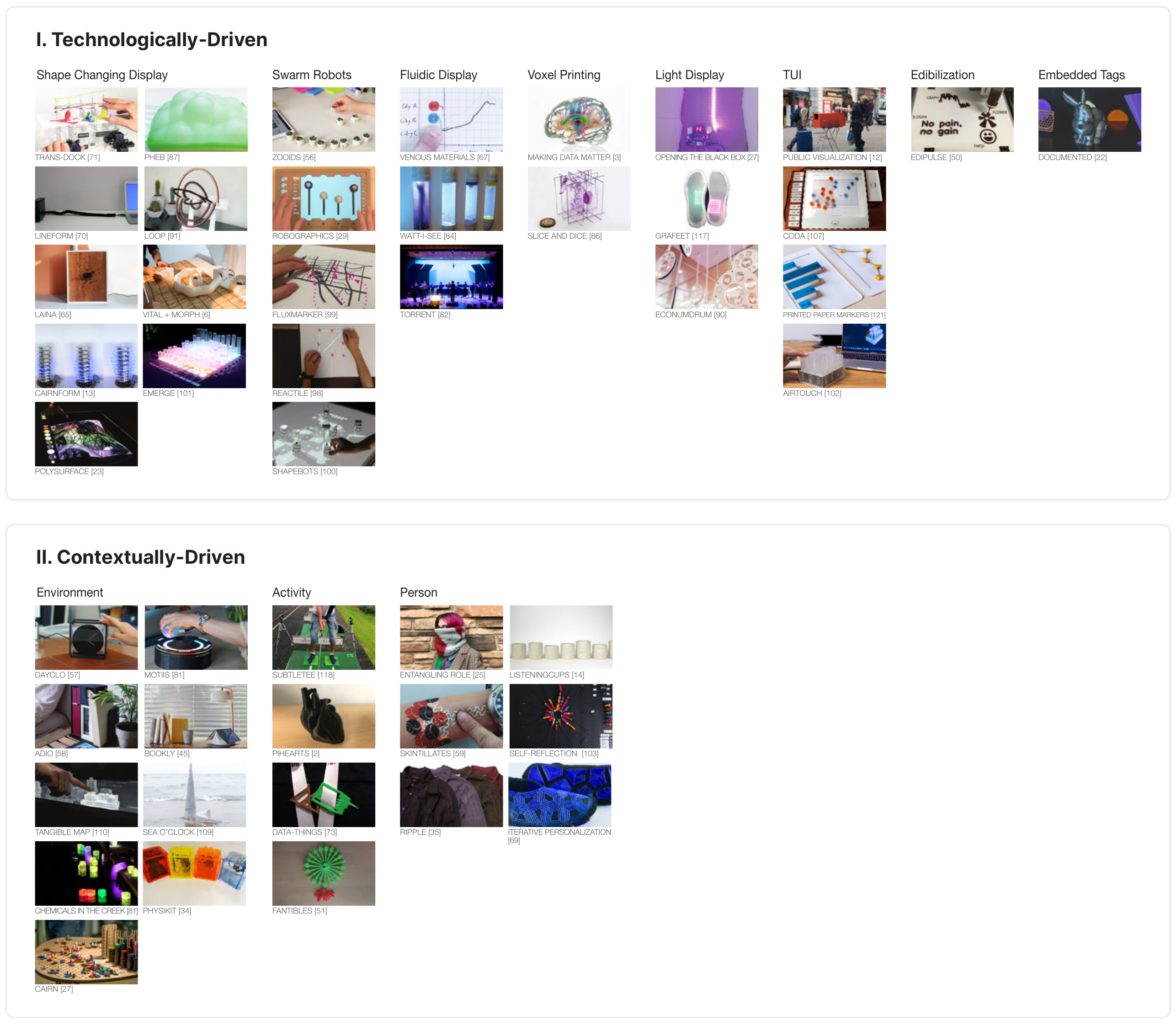}
  \caption{Eleven examples of physicalization embodiments identified in our corpus. The physical embodiment is categorized as either technologically- or contextually driven. All images are copyright to their respective owners.}
  \label{fig:embodiment}
  \Description{Examples of physicalization embodiment. This picture has 2 big boxes. The Left box A Technology-Driven contains 8 categories. Shape-Changing Display has 9 project images. Swarm Robotics has 5 project images. TUI has 4 project images. Fluidic Display has 3 project images. Light has 3 project images. Voxel Prints has 2 project images. Embedded Tag has 1 project image. Edibilization has 1 project image. The Right box B Context Driven has 3 categories. Environment has 9 project images below it. Person has 6 project images. Activity has 4 project images.}
\end{figure*}

\subsection{What design decisions contribute to a physicalization's materiality?}
\label{sec:design-decisions}
Past work has analyzed physical representation in physicalization through a number of lenses, including pragmatic/artistic \cite{hogan2012multimodal, djavaherpour2021rendering}, epistemological/ontological \cite{offenhuber2021physicality}, and semiotic perspectives \cite{zhao-embodiment2008, offenhuber2015indexical}. To understand physicality within our design space, we investigated each artifacts' materiality and found three qualities of existence specific to physicalization: embodiment, data duration, and material (Figure~\ref{fig:materiality}). We developed a high-level categorization of physicalizations as either technologically-driven or contextually-driven by specifically focusing on how data is \textit{manifested} physically and temporally. We identified 11 types of embodiment (Figure~\ref{fig:embodiment}), three classifications of data duration (described in Section \ref{sec:space_structure}), and distinguish between materials as either electronic or non-electronic.

\textbf{Embodiment}
A physicalization's embodiment provides insight into the designer's domain expertise, motivations, techniques, and encoding schemes. 
We categorized embodiment based on how each physicalization manifested data and whether they were  \textit{technologically-driven} or \textit{contextually-driven}. 
Technologically-driven physicalizations consisted of platforms where designers leveraged technology to encode data. These platforms included shape-changing displays, swarm robotics, light displays, TUIs, fluidic displays, edibilization, vibrotactile displays, embeddable tags, and voxel prints. Within shape-changing displays, technologies ranged from inflatables to programmable pin arrays to actuated hand-held devices. The materials for technologically-driven physicalizations were chosen based on their performance or use within a particular manufacturing method (e.g., 3D printing, laser cutting, silicone molding). 

Physicalizations whose embodiment was contextually-driven represented data through people, environments, or activities. In these cases, physicalizations heavily relied on the context surrounding its use or the user themselves \cite{lo2016skintillates, desjardins2019Listening, howell2018tensions}. These systems emphasized data mapping choices similar to traditional visualization.

\textbf{Material}
We categorized the materials used to construct a physicalization as \textit{electronic} or \textit{non-electronic}.  We intentionally do not use the more common HCI practice of describing an object as either programmable or non-programmable because non-electronic materials can be transformed through craft and other manipulations to become programmable (e.g., the Embroidered Computer \cite{kurbak2018stitching}). We characterize a material as \textit{electronic} if any manufactured component or collection of manufactured components are used to \textit{control} the flow of current (microcontrollers, sensors, computers, active and passive components, etc.). 
Physicalizations that used electronic components can generally be reused with different datasets for new lines of inquiry and analysis (e.g., \cite{le2018dynamic, nakagaki2020transdock, everitt2017polysurface, taher2016emerge}). 
Physicalizations that had no computational abilities typically targeted a single dataset (e.g., \cite{thudt2018self, friske2020data, vonompteda2019sea, mor2020venous}) or required manual reconstruction of the physicalization by the user.

\begin{figure}[b]
  \centering
  \includegraphics[width=\columnwidth]{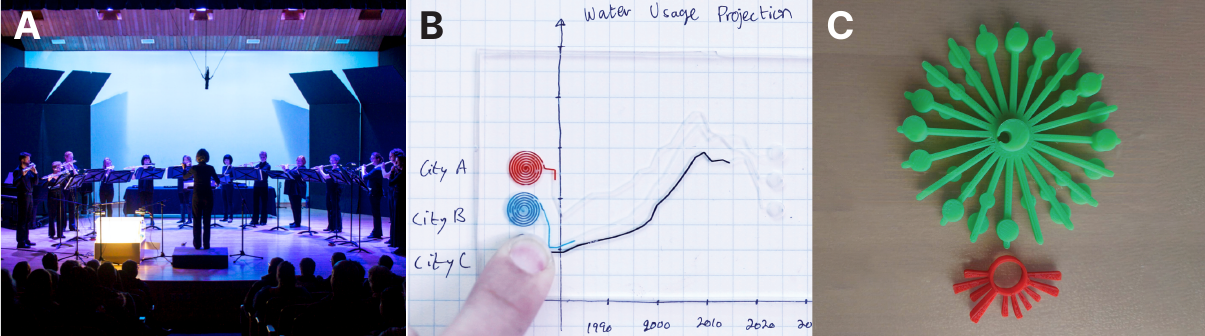}
  \caption{Examples of physicalizations with different data duration (i.e., how long data is displayed through its physical embodiment): (a) Torrent \cite{pon2017torrent} displays data ephemerally through its use of water, (b) Venous Materials \cite{mor2020venous} displays data persistently as determined by the user, (c) Data is permanently embedded in Fantibles \cite{khot2016fantibles}.}.
  \label{fig:time}
  \Description{Three examples of different data duration in physicalization. Picture A shows 12 flutists with a conductor, and a lighted-up water box next to her on stage in front of an audience. Picture B shows two axes. Vertical line pointing up with hand-written text of Water Usage Projection. The bottom text below the horizontal line shows 1990, 2000, 2010, 2020. The left side of the vertical axis has 3 labels, City A, City B and City C, corresponding to 3 colored coiled fluid channels. An index finger is pressing the coil of City C and the colored fluid line shows a line graph curving up from 1990 toward 2010 and curving down toward 2020. Picture C shows 2 plastic spiky objects with radial patterns. The top green spike is named SPORTOMS that has 20 spikes. The text Start of the game is pointed to the hole in the middle of this spike. Each spike = 5 overs. The small circles on the spikes represent Wickets. The bigger circles at the end of the spikes are Runs. The bottom red object is called FANTOMS. It has a circle with the lengths of spikes representing emotions in Low (no spike), Medium (medium length) and High (longer length).}
\end{figure}

\textbf{Data Duration}. Physicalizations are often designed for one of three temporal durations: \textit{ephemeral}, \textit{persistent}, and \textit{permanent} (Figure~\ref{fig:time}). Physicalizations whose data is presented \textit{ephemerally} collect and render data in real-time and cannot be recorded or recalled. We found that all of these physicalizations (12/\corpusNum) use sensors to collect and visually stream data. For example, Torrent \cite{pon2017torrent} (Figure~\ref{fig:time}a) 
augments concert flute music with live water sounds. The water surface turbulence reflects the flutist' muscle tension throughout the performance. Physicalizations that \textit{persist} offer users some control over how long data is displayed. These physicalizations 
often rely on electronics and programming to specify how to retain the data embedded in the physicalization. However, other systems use manual interactions to control data persistence. Venous Materials~\cite{mor2020venous} (Figure~\ref{fig:time}b)  uses microfluidics for tangible interactions to represent a dynamic line chart. By directly pressing on a chamber containing the liquid, a user can control the flow of liquid through the microfluidic channel with their finger. People decide how long they would like to see data being displayed before releasing their finger and returning the display to its original state. Physicalizations that display their data \textit{permanently} are those where the data is permanently embedded or integrated into its physical structure, such as conventional 3D printed objects \cite{khot2016fantibles, nissen2015data, bader2018making, desjardins2019Listening} or hand-crafted artifacts \cite{friske2020data, thudt2018self}. 
Most persistent artifacts did not include any electronic materials. As noted above, such physicalizations cannot decouple materials from data \cite{friske2020data, bader2018making}, making it difficult to repurpose the artifact for a different dataset without significant effort from the user.

\subsection{What interactions does a physicalization afford?}
\label{sec:affordances}
\begin{figure}[t]
  \centering
  \includegraphics[width=\columnwidth]{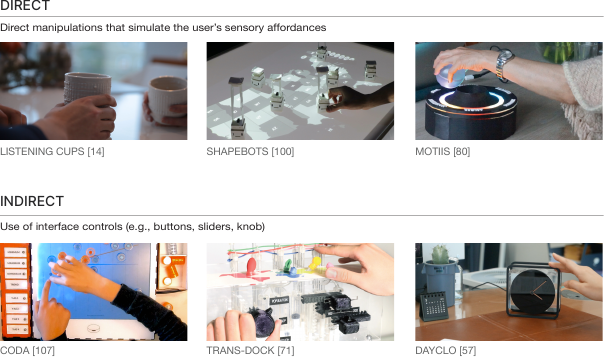}
  \caption{Examples of physicalizations using the \textit{explore} interaction. Physicalizations using the \textit{direct} mechanism uses direct manipulation of the object with a focus on sensory affordances. \textit{Indirect} refers use interface controls (e.g., buttons, sliders, knob) to interact with the data. FluxMarkers \cite{suzuki2017flux} uses speech and touch to explore the data.}
  \label{fig:explore}
  \Description{Examples of Explore Interaction for physicalization. Direct examples show 3 project pictures: a hand holding a cup, a hand grabbing a small robot on a projected US map, and a hand holding a ball on top of a disk shape. The Indirect examples show 3 project pictures: two hands touching multiple coiled liquid channels: the left hand holding a dial next to the label of population, while the right hand touching one of the different colored balloons controlled by mechanisms with multiple tubes, and a hand touching the corner of the box containing a clock.}
\end{figure}

As discussed in Section~\ref{sec:interactions-codebook}, we analyzed how the physicalization's design features (i) invite people to conduct a particular data interaction (i.e., perceived affordances), (ii) informs what the result of their performed data interaction will be (i.e., feedforward), and (iii) communicate the result of their action (i.e., feedback).
\textit{Explore} (i.e., to interrogate different subsets of the data~\cite{yi2007interactions}) was the most pervasive interaction technique (12/61). 
We saw \textit{explore} applied to everyday objects (e.g., clocks~\cite{lee2020dayclo}  cups~\cite{desjardins2019Listening}) and shape-changing displays~\cite{suzuki2019shapebots, nakagaki2020transdock}. Systems that supported \textit{explore} allowed people to explore data either directly (akin to direct manipulation in visualization) or indirectly (i.e., akin to dynamic queries \cite{shneiderman1994dynamic}; Figure~\ref{fig:explore}).

Physicalizations that used direct interactions stimulated the user's sensory affordances \cite{hartson2003affordances}. Most such physicalizations were composed of discrete pieces where each piece is part of the system's overall interaction scheme. 
For example, ShapeBots \cite{suzuki2019shapebots} (Figure~\ref{fig:explore}) is a swarm robotics system where each robot is encoded as a data point within a dataset. With ShapeBots, the researchers intentionally designed the robots to be small, such that the graspable, small pieces invite the user to pick up each piece and interact with the physicalization as a system of parts. ListeningCup~\cite{desjardins2019Listening} also used form to encourage interaction, inviting people to stroke the cup with their hands through its textured surface. In both cases, these systems afford feedforward by primarily appealing to the perceptual-motor skills. ShapeBots' small graspable components inform people that data can be picked up, and the texture on ListeningCup offers visual cues as to what it might \textit{feel} like.\footnote{For more discussion on sensory-driven design, we refer readers to Hara \cite{hara2007designing}.} With the exception of ListeningCups, all physicalizations with direct interactions were made with electronics that responded immediately, changing their state to serve as feedback. For instance, moving a ShapeBot \cite{suzuki2019shapebots} from one state to another state evoked a change in height, moving a Physikit cube \cite{houben2015physikit} from one part of the house to another created change in sensor readings and output in the form of light, vibration, airflow, and movement. ListeningCups, in contrast, provided feedback through its texture which could be experienced through direct touch. 

Physicalizations that used indirect interaction in contrast presented physical controls that people are familiar with (e.g., buttons, sliders, knobs; Figure~\ref{fig:explore}). People were informed of what the interactive component (e.g., the button or slider) would do through lexical or graphical labels on the physical controls.
Similar to direct mechanisms, these physicalizations communicated the result of their action by immediately changing the state of physicalization through use of electronic components.

As physical artifacts, data physicalizations leverage our abilities to manipulate everyday tangible things (e.g., cups, blocks, lamps, clocks) to make sense of intangible data. In addition, data physicalizations with computational components may leverage interface conventions (e.g. sliders, knobs, buttons) to facilitate interaction. Echoing Jansen et al. \cite{jansen2015opportunities}, we observe many parallels between data physicalizations and TUIs. 
Examining the affordances of physicalizations might enable designers to better leverage TUI principles.

The aforementioned physicalizations employ interaction techniques reflecting known concepts from TUIs, such as physical interactions through graspable elements \cite{Fitzmaurice1996graspable} and token+constraint systems \cite{ullmer2005tokenconstrained}.
However, interactions within data physicalizations in our corpus---like tangible interactions---cover a wide spectrum of input techniques, from indirect physical controls to more direct coupling of tangible interfaces with digital data (e.g. \cite{suzuki2018reactile, guinness2019robographics, veldhuis2020coda}, see ``tangible bits'' \cite{ishii1997tangiblebits}) and even pushing towards programmable physical matter (e.g. \cite{nakagaki2020transdock, mor2020venous, legoc2016zooids}, see ``radical atoms''~\cite{ishii2012radical}).

With this breadth of different tangible interaction techniques and concepts available, we reiterate that---as interfaces---affordances remain a fundamental consideration while designing physicalizations~\cite{norman1988psychology}. Affordances for data physicalizations bridge considerations that tie \textit{context}, \textit{interactions}, and \textit{structure} together.
For instance, connecting the choice of material and embodiment to interactions can help tailor a physicalization to a particular target audience. FluxMarkers \cite{suzuki2017flux} and Robographics \cite{guinness2019robographics} connect electronic swarm robots to the BLV community. These systems both incorporate sound as either input or output and vibrotactile haptic feedback for multi-sensory interaction. 

\subsection{How do different communities approach physicalization? (And what can we learn from each other?)}
\label{sec:community}
Communities involved in physicalization share many of the same values, such as the importance of embodiment and designing specific user tasks. The visualization community has illustrated the importance of understanding a user's task. 
A visualization whose focus is to reveal analytical insight for domain experts will be radically different compared to others that depict data for casual use for the general public \cite{Pousman2007casual}. We contend this logic also extends to physicalizations, and we see an opportunity for different communities to learn from each other by understanding how to best express data with respect to a given task. Physicalizations in TUI and design more deeply explored the value of reflection and agency in design, offering insight into how physicalization might promote self-awareness and prolonged engagement. 

\subsubsection{Analysis: Importance of Effectiveness}
Physicalization is often used for problem-solving and decision-making (i.e., analysis) \cite{jansen2015opportunities}. \textit{Analysis} is the second most pervasive task (15/\corpusNum), and other communities can learn from the visualization community specifically in learning how to design systems based on visual effectiveness (e.g., graphical perception) and system flexibility. 

As mentioned in Section~\ref{sec:definition}, Jansen et al. \cite{jansen2015opportunities} laid out a research agenda for empirical work that can lead to a better understanding of the effectiveness of physical variables. Ignoring this groundwork may lead to ineffective designs where people are not able to perceive changes in data through the transformations presented by the physicalization. We see this reflected within our corpus where researchers have used \textit{volume} to communicate changes. Two such examples include Bookly \cite{ju2019bookly} and CairnFORM \cite{daniel2019cairn} which both encode data using volume. CairnFORM consists of a tower of ten actuatable disks which expanded and retracted in the XY plane. Participants in their user study expressed difficulty in discerning the changes in the diameter of each disk. Bookly is designed with a single motorized disk that expands in the z-direction. Most of their participants did not notice the volumetric change in real-time. Both works discussed how users of their respective systems found it difficult to perceive changes in volume. However, studies in visualization show that people have trouble reasoning over changes in volume and are better at reasoning over other visual variables such as surface area \cite{jansen2015psychophysical}. Work such as \cite{sauve2021reconfiguration, jansen2015psychophysical, garcia2021scale} can help researchers design systems that utilize effective physical variables. 

Work from the visualization community also highlights the importance of allowing systems to adapt to a given user's needs. Polysurface \cite{everitt2017polysurface} provides such flexibility through the actual visual design of their system:
the shape-changing display was configurable to three different domain experts' specific needs. In contrast, Zooids \cite{le2018dynamic} and EMERGE \cite{taher2016emerge} provided flexibility through data interactions. As mentioned in Section~\ref{sec:audience}, the depth of the supported data analysis varies depending on the number of interactions available.

\subsubsection{Reflection: Importance of Agency}
\label{sec:agency}
Physicalizations designed for reflective tasks, in contrast, highlight the importance of agency---users should be able to develop personally meaningful data mappings. These customized physicalizations tend to layer meaning-making by encouraging the user to explore and reflect as part of the design process as well as through interacting with the physicalization. A notable example comes from a user within Thudt et al.'s study \cite{thudt2018self} where tokens were used to visually distil a recipe for an ``at a glance'' overview. The beads used color to encode ingredients while the number of beads of a particular color encoded how much of each ingredient to use relative to one another.  The mapping process inadvertently created a label for the jars of the product generated by each recipe. 

Not all physicalizations for reflection provided agency to the users. Data Things \cite{nissen2015data} and Fantibles \cite{khot2016fantibles} leveraged users' personal data such as social media posts but did not include the users in the data mapping process. This exclusion led people to alter their social media feed to reshape their physicalizations and regain their sense of agency during participatory workshops. Even though the design and encoding processes did not intentionally encourage agency, people altered their own behavior to influence the outcome of the physicalization's form. 

\section{Discussion}
\label{sec:discussion}
Physicalization draws on traditions and practices from a number of disciplines, including data visualization, TUI, and design. Our survey aligns recent developments across these fields to establish a preliminary design space of 13 key dimensions for physicalization design and development. In developing this design space and synthesizing trends and patterns in physicalization research, we first identify two priorities researchers and designers should consider (Section~\ref{sec:consider-context} and Section~\ref{sec:define-data}) and offer opportunities for future work and interdisciplinary collaboration (Section~\ref{sec:synthesis-use}). 

\subsection{Considering Context}
\label{sec:consider-context}
Context grounds the visual design and functionality behind a data physicalization. We found several instances where the designers of physicalizations were not explicitly clear about the physicalizaton's context, specifically its target audience, location, and/or task (Section~\ref{sec:design-space-context}). While we cannot assert that context was not considered in the design of these artifacts, context is key for both creating effective representations and for connecting relevant bodies of research across disciplines. For example, the tasks people need to perform with their data guide many visualization design workflows and ensure the resulting physicalization communicate attributes of the data that matter to the individual \cite{sedlmair2012design}. 
Emphasizing context can help researchers identify further end-user considerations that might have not been captured by current research. 

Context also offers design guides that may help strengthen physicalizations. For example, physicalizations that aim to support analysis should consider \textit{data duration}---generally reducing ephemerality and aiming for more persistent or permanent designs to enable people to return to key states that elicit insight into their data (Section~\ref{sec:design-decisions}). Designers may also draw on concepts from data provenance to inspire new methods to physicalize data that allow analysts to retrace states of the physicalization over time during analysis \cite{ragan2015characterizing}.

\subsection{Defining Data}
\label{sec:define-data}
We used Jansen et al.'s  and Hogan and Hornecker's definitions \cite{jansen2015opportunities, hogan2012multimodal} of physicalization to construct our corpus. However, there is not a clear consensus on what constitutes a physicalization and, more generally, what constitutes data. 
Offenhuber \cite{offenhuber2021physicality} notes that there are several definitions of data that vary across bodies of scholarship, including data that stems from the mind (i.e., epistemological). This perspective contrasts the traditional notion of data in the visualization community (i.e., ontological). 
In works such as Data Agents~\cite{Karyda2021data} and Imagining Movie Recommendation Algorithms~\cite{alvarado2021netflix}, mundane objects were used to represent people's personal data in meaningful ways filtered through the creator's internal synthesis of the data. These systems do not meet Jansen et al.'s and Hogan and Hornecker's definitions as data was not physically represented within the physical and visual properties of these data objects or their behaviors. Instead, the artifacts reflect an internal synthesis of data, creating objects whose form is indirectly shaped by data rather than directly representing individual data points.  

The reflective focus of many physicalizations and their roots in design and artistic practice challenge traditional definitions of data and data representation from communities like IEEE VIS. 
One concrete example is the difference in the underlying views of data between the two communities. VIS often takes \textit{positivist} approach, which views knowledge and data as objectively determined \cite{meyer2019rigourdesign}. In contrast, the design community frequently employs \textit{interpretivist} approach where knowledge is viewed as socially constructed and reflected upon.
Interpretivist physicalizations consider not only the data's form, but also how that form may introduce social dynamics that shape interpretation and interaction with the physicalization (e.g., Friske et al. \cite{friske2020data}, Section~\ref{sec:audience}). People may see the physicalization as ``correct'' and objective in ways that may draw into question competing interpretations of the same dataset. However, rather than invalidate a given interpretation and narrative drawn from physicalized data, we call the communities to consider a broadened definition of data and how we approach, consider, and render it.

\subsection{Synthesis and Use}
\label{sec:synthesis-use}
Through our design space, we observed multiple dimensions associated with the design of data physicalization.
Physicalization designers can use this design space to reason across relevant bodies of research, identify new collaborative opportunities, and explore methods for creating and evaluating physicalizations. 

A primary motivation of this paper is to bridge knowledge from visualization, TUI, and design in order to provide a cohesive lens of each relevant intellectual community's set of approaches, theories, and terminologies. 
The presented design space is intended to help designers and researchers be more cognizant of knowledge and practices emerging in other communities (Section~\ref{sec:community}). This need is especially pertinent as our conducted survey shows data physicalization researchers come from various backgrounds (as seen in Section~\ref{sec:summary}), yet these papers seldom draw substantively on the literature of the other communities. While the lack of cross-referencing does not itself imply a lack of interchange, we also found a lack of cross-pollination of ideas, techniques, and methods. Our survey aims to bridge by presenting a shared context for exploring vital facets of physicalization design, including the value of physicalizations as means for for data communication in visualization, for expressivity and reflection in design, and for understanding and demonstrating novel techniques and material use in TUI.

The interdisciplinary nature of data physicalization also opens new opportunities for closer collaborations among relevant communities. As shown in Section~\ref{sec:community}, we argue for closer collaboration particularly from the standpoint of how cross-pollination may aid in \textit{generating} new data physicalization artifacts and \textit{evaluating} existing ones. This interdisciplinary cross-over can happen at both a low-level (e.g., integrating data visualization knowledge regarding visual perception) and at a higher-level (e.g., considering how data is integrated into a physicalization's overall form).  Future work generating and evaluating data physicalization may benefit from a research through design approach, borrowing on related methods employed in each of the surveyed fields.

\subsubsection{Research through Design}
Research through design is prevalent in visualization \cite{sedlmair2012design, meyer2019rigourdesign} (often instantiated through design studies) and the broader HCI community \cite{2007zimmerman} because of its emphasis on synthesizing knowledge into meaningful artifacts for both research and practice. Each data physicalization artifact is (un)consciously a product of different community perspectives and values. Collaborative applications of research through design in physicalization offer an opportunity to investigate new techniques through the lens provided by each community. However, successful applications of these methods and integration of community perspectives requires considering how we can be more intentional with the physicalizations we produce 
to enable the blending of ideas and values. 
We encourage designers and researchers to refer to the criteria laid out by Zimmerman et al.~\cite{2007zimmerman} and Meyer and Dykes~\cite{meyer2019rigourdesign} for integrating rigorous design into their research practice. These frameworks emphasize synthesizing diverse considerations for constructing and evaluating research artifacts in the real world. 
From our analysis of the data physicalization corpus covered in this paper, two criteria stand out to us in particular:

\textbf{Invention.} One example of generating new design possibilities 
is through the lens of \textit{invention} \cite{2007zimmerman}.
For example, a multidisiplinary lens on interactive physicalization may offer innovations 
for more sustainable interaction design. Section~\ref{sec:design-decisions} and Section~\ref{sec:interactions} show that most electronics physicalizations provide easy updates to data representations due to the automation and computation capabilities offered by electronic devices. These properties make the physicalization reusable with multiple datasets and often increase their data and task scalability. However, electronic physicalizations are often expensive, complex, and ecologically unsustainable \cite{everitt2017polysurface}. 
Our survey shows that electronic materials are not the only technology that can provide automation and computation. Mechanical, chemical, and biological actuated physicalization offer many of the same computational benefits of electronic physicalization while potentially being more ecologically friendly. 
For example, Venous Materials \cite{mor2020venous} provide interaction based on microfludiics, and Printed Paper Markers \cite{zheng2020printed} employ kirigami techniques for actuation design where paper acts as buttons with tactile feedback. Future work in data physicalization can consider new or emerging domains relevant to HCI, such as the physical and biological sciences, to investigate new interaction paradigms and design dimensions.

\textbf{Extensibility/Resonance.} Both Zimmerman et al. \cite{2007zimmerman} and Meyer and Dykes \cite{meyer2019rigourdesign} contend that the artifacts produced through design studies should communicate implicit, hidden knowledge and build upon existing knowledge. Applying this perspective across multiple communities may help designers better comprehend and leverage embodiment in physicalization. As discussed in Section~\ref{sec:design-decisions}, we identified 11 types of physicalization embodiment. This variety indicates a growing exploration of how to materialize data. However, there is a lack of knowledge transfer among each embodiment type (i.e., technology-driven, contextually-driven) and category (e.g., shape-changing display, swarm robotics; Figure~\ref{fig:embodiment}). Throughout our analysis, we found that researchers seldom explicitly situate their artifact within the bigger landscape of data physicalization. This approach can silo physicalizations from each other, limiting our abilities to connect between different innovations and subdisciplines. We anticipate that our design space can help remedy this siloing
by providing a shared physical representation vocabulary. 
Moving forward, we encourage researchers and designers to use this design space as a starting point to draw inspiration for new physicalizations and transfer knowledge produced from their artifact to general data physicalization knowledge.

Collectively, our work suggests the need for the data visualization, TUI, and design communities to engage in a collaborative dialogue around design motivations, methods, and techniques for physicalization development. Such cross-disciplinary collaboration will lead to mutually beneficial innovation that will help advance the field of data physicalization.
We hope this paper serves as a call to action for members of these communities to recognize and take advantage of potential synergies among the relevant bodies of knowledge to drive novel approaches to physicalization.

 \subsection{Limitations}
Grounded theory requires data collection in an environment constructed by the researchers. Although measures were put in place to maximize credibility and repeatability, code application depends on the definitions and perspectives of the coders. Different coders may result in different findings. Papers within the corpus were not always explicit with respect to each dimension and were sometimes non-exhaustive (e.g., locations or audiences were not always described), introducing additional uncertainty. To mitigate potential bias, the authors
consist of researchers from various fields, and we used peer debriefing to ensure that conclusions were grounded in the data. 
Physicalizations are diverse not only in their structure but use and interpretation of data (Section~\ref{sec:define-data}). This diversity created challenges in determining whether a paper belonged within the corpus. 
While we aim for coverage of a range of HCI disciplines, our survey is not intended to be comprehensive. 
Physicalizations exist beyond research articles (e.g., \cite{segal, tali}), which we have not covered within our corpus.
This choice limits the scope of our survey to artifacts with an emphasis on human interactions with data as a product of technological innovation (e.g., materials engineering, electronic interactions, data analytics, craft). Communities of practice outside of HCI, such as art, may offer alternative frameworks for reasoning about physicalization design and complement the dimensions of our design space. Future work should investigate and refine the design space through the integration of a broader range of knowledge.

\section{Conclusion}
\label{sec:conclusion}
We surveyed \corpusNum\space physicalizations to understand common design elements of physicalization research across visualization and HCI. We used these elements to characterize thirteen dimensions of physicalization work. Our survey revealed that physicalizations offer solutions for a diverse range of tasks with objectives and designs that extend beyond the analytical, expert-driven focus of systems from conventional visualization research, such as emphasizing reflection and public engagement. These innovations arise from the deeply cross-disciplinary traditions of the field. We reflect on these traditions and gaps in the design space to identify key open challenges in physicalization research that require rich, collaborative perspectives to address. We hope that this perspective can proactively bridge the diverse knowledge and techniques that physicalization research inspires.   
\begin{acks}
This research is sponsored in part by the U.S. National Science Foundation through grants IIS-2040489 \& STEM+C 1933915. 
\end{acks}

\bibliographystyle{ACM-Reference-Format}
\bibliography{bib}

\end{document}